\documentclass[preprint,12pt]{elsarticle}
\usepackage{subfigure}
\usepackage{amsmath}
\usepackage{amssymb}
\usepackage{url}
\biboptions{sort&compress}
\usepackage[a4paper,total={6in, 8in}]{geometry}
\usepackage{atbegshi}
\AtBeginDocument{\AtBeginShipoutNext{\AtBeginShipoutDiscard}}

\journal{Physics Letters A}

\DeclareMathOperator{\sign}{sign}

\begin{document}

\begin{frontmatter}

\title{Towards skyrmion crystal stabilization in the antiferromagnetic triangular lattice at ambient conditions}
\author[1]{Mariia Mohylna},
\author[1]{Vitalii Tkachenko},
\author[1]{Milan \v{Z}ukovi\v{c}\corref{cor1}}
\ead{milan.zukovic@upjs.sk}
\address[1]{Department of Theoretical Physics and Astrophysics, Institute of Physics, Faculty of Science, Pavol Jozef \v{S}af\'arik University in Ko\v{s}ice, Park Angelinum 9, 041 54 Ko\v{s}ice, Slovak Republic}
\cortext[cor1]{Corresponding author}


\begin{abstract}
We explore possibilities of stabilizing a skyrmion (SkX) phase in the frustrated antiferromagnetic (AF) Heisenberg model with DMI on a triangular lattice at ambient conditions. We show that by stacking multiple AF layers and coupling them with strong ferromagnetic (F) interaction the SkX phase stability temperature window can be extended by several times. The transition temperature increase is supported by the increasing number of layers and interlayer interaction and thus room-temperature SkX phase could be achievable by proper tailoring of the multilayer structure. Further, we demonstrate that in the sandwich heterostructure, obtained by inserting a stiff F layer in between two AF layers, the required external magnetic field can be completely replaced by the effective field imposed by the F layer. Thus, one can design a heterostructure with the SkX phase stabilized at relatively high temperatures and zero field - the environment favorable for technological applications.
\end{abstract}



\end{frontmatter}

\section{Introduction}

Magnetic skyrmions are topological spin textures that are stabilized in various types of magnets by different kinds of interactions, most commonly by the Dzyaloshinskii–Moriya interaction (DMI)~\cite{dzyaloshinsky1958thermodynamic, moriya1960anisotropic}. After their experimental observation in 2009~\cite{muhlbauer2009skyrmion} most theoretical and experimental works focused on finding the skyrmion lattice (SkX) phase or individual skyrmions in ferromagnets (F) \cite{belavin1975metastable, bogdanov1989thermodynamically, bogdanov1994thermodynamically,muhlbauer2009skyrmion, do2009skyrmions, heinze2011spontaneous}. Nevertheless, more recently the attention has been shifted to antiferromagnetic (AF) materials. The rich properties of AFs, such as zero net magnetization, tunable AF order, tunable spin-orbit torque strength, and large exchange bias, combined with higher robustness and elimination of undesirable effects in the skyrmion current~\cite{zhang2016antiferromagnetic, barker2016static, bessarab2019stability}, provide many future opportunities to explore and optimize skyrmion-based applications. Despite intensive theoretical investigations, the first AF skyrmion-like textures have been experimentally realized only very recently~\cite{gao2020fractional,jani2021antiferromagnetic}.

Several theoretical studies focused on the formation and stability of AF skyrmion crystal (AF-SkX) state, consisting of three interpenetrating skyrmion crystals, observed in the frustrated AF triangular Heisenberg magnet with the presence of DMI~\cite{rosales2015three, osorio2017composite, mohylna2021formation, mohylna2022stability, fang2021spirals}. In spite of the observation that a uniaxial magnetic anisotropy can strongly affect spin ordering in the frustrated triangular magnet~\cite{leonov2015multiply}, it has been shown, that the AF-SkX phase can be stabilized in a quite wide temperature-field window within relatively broad ranges of the DMI and the anisotropy strengths. However, magnetic skyrmions in the AF-SkX phase can only be observed in a rather strong out-of-plane magnetic field, which may limit their use in magnetic memory technologies or even inhibit their observation in the experiment. For example, in Fe/MoS\textsubscript{2} - the potential experimental realization of the SkX phase in the present model - the required magnetic field is estimated to be as high as 20 T and it only appears at rather low temperatures~\cite{fang2021spirals}. Thus, significant reduction or complete elimination of the external magnetic field needed to stabilize the SkX phase and extending the temperature window of the skyrmions' stability to ambient temperatures represents an important step towards the application of skyrmions in high-density magnetic memories, data storage, and computing devices.

It has been suggested that the stabilization of the skyrmion state at very low or zero external fields and room temperatures can be achieved by designing heterostructures composed of bilayers or multilayers~\cite{chen2015room,yu2018room, guang2020creating, nandy2016interlayer,brandao2021stabilizing}. For example, Chen~\emph{et~al.} showed an approach to stabilize a room temperature skyrmion ground state in chiral magnetic films via exchange coupling across non-magnetic spacer layers~\cite{chen2015room}. Another example of the room-temperature skyrmion phase stabilization is the system based on AF and F metals IrMn/CoFeB heterostructures~\cite{yu2018room}. In an exchange-biased magnetic multilayer individual skyrmions at zero field have been created by exposure to soft X-rays~\cite{guang2020creating}. Some other recent mechanisms to create zero-field skyrmions at room temperature in ferromagnetic multilayers are reviewed in Ref.~\cite{brandao2021stabilizing}. 

An elegant yet rather simple approach to the significant reduction of the giant field required for the presence of skyrmions in the thin chiral magnet is based on its coupling to a hard F layer with strong out-of-plane anisotropy~\cite{nandy2016interlayer}.
Then, the F-ordered reference layer via the interlayer exchange coupling induces the effective field that collaborates with or even fully substitutes the external magnetic field in the skyrmion formation. This only happens if the reference layer is a stiff ferromagnet with a strong out-of-plane anisotropy and the effective field is in the range of the magnetic field required for skyrmion stabilization. Such conditions can be created by varying the thickness of the layers and the composition at the interface, resulting in the effective fields exceeding 40 T in the Mn/W\textsubscript{m}/Co\textsubscript{n}/Pt/W(001) multilayer system with 5 layers of W. 

To achieve skyrmions at ambient conditions, perpendicularly magnetized multilayers made of nano-thick magnetic and non-magnetic materials have been used as a platform with suitable parameters to host skyrmions at room temperature~\cite{moreau2016additive}. Multilayer stacks composed of multiple repetitions of thin magnetic metal layers separated by heavy metal nonmagnetic layers grown by sputtering deposition enable the increase of the thermal stability of columnar skyrmions, that are coupled in the successive layers~\cite{back20202020}.


Most of the studies focused on magnetic multilayers involving F metal and heavy metal layers. However, incorporating AFs in magnetic multilayers can lead to a number of important technical advances in spintronics such as current-driven zero-field magnetization switching and ultrafast control of spins in adjacent F layers. Designing of such heterostructures can take advantage of tunable interfacial interactions at the AF/F interface including perpendicular magnetic anisotropy, exchange
bias, and DMI, as well as the optimization of these coexisting interactions. Thus, searching for skyrmions in multilayer heterostructures incorporating AF layers may provide new opportunities to further improve device performance~\cite{yu2018room}.

In the present Letter, we propose a mechanism for stabilizing the AF-SkX phase in the frustrated AF triangular Heisenberg magnet with DMI by designing suitable multilayers/heterostructures. In particular, we demonstrate that by stacking 2D layers in the vertical direction, one can considerably enhance the thermal stability of the skyrmions and thus extend the temperature window of their persistence by several times. Furthermore, by inserting a suitable reference F layer in between the AF layers the AF-SkX phase can be stabilized at much smaller external fields, including zero field.

\section{Models and Method}

In this study we consider two structures that include 2D layers of the Heisenberg AF triangular lattice with DMI, capable of hosting the AF-SkX phase. The first one (AF multilayer) consists of $n_l$ AF layers vertically stacked on top of each other, described by the Hamiltonian 
\begin{equation}
\begin{split}
\mathcal{H_L} = & - J_{\parallel} \sum_{\langle i,j \rangle}\vec{S_{i}}\cdot
\vec{S_{j}} + \sum_{\langle i,j \rangle}\vec{D_{ij}}\cdot\Big [\vec{S_{i}}\times\vec{S_{j}} \Big] \\
& - J_{\perp} \sum_{\langle i,k \rangle}\vec{S_{i}}\cdot\vec{S_{k}} - h\sum_i S_i^{z},
\end{split}
\label{Hamilt_multi}
\end{equation}
where spins $\vec{S_i}$ and $\vec{S_j}$ belong to the same and $\vec{S_{k}}$ to the adjacent layers. The first three summations run over all nearest neighbor (nn) pairs and the last one over all spins in the system. The intralayer coupling $J_{\parallel}<0$ is AF, the interlayer one $J_{\perp}>0$ is F, the DMI vector $\vec{D}_{ij}=D\vec{r}_{ij}/|\vec{r}_{ij}|$ is oriented along the radius vector between two neighboring sites $i$ and $j$, and $h$ is the external magnetic field applied along the stacking direction.


The second heterostructure (AF/F/AF sandwich) consists of two AF layers, coupled via next-nearest-neighbor (nnn) interactions, and one F layer with out-of-plane anisotropy sandwiched in between. The Hamiltonian reads

\begin{equation}
\begin{split}
\mathcal{H_S} = & - J_{AF} \sum_{\langle i,j \rangle}\vec{S_{i}}\cdot
\vec{S_{j}} + \sum_{\langle i,j \rangle}\vec{D_{ij}}\cdot\Big [\vec{S_{i}}\times\vec{S_{j}} \Big] \\
& - J_{F} \sum_{\langle \alpha,\beta \rangle}\vec{S_{\alpha}}\cdot\vec{S_{\beta}} - A\sum_{\alpha} (S_{\alpha}^{z})^2 \\
& - J_{AF-F} \sum_{\langle i,\alpha \rangle}\vec{S_{i}}\cdot\vec{S_{\alpha}} - J_{AF-AF} \sum_{\langle k,l \rangle}\vec{S_{k}}\cdot\vec{S_{l}} - h\sum_i S_i^{z},
\end{split}
\label{Hamilt_sand}
\end{equation}
where we consider the exchange nn interactions within the AF planes ($J_{AF}<0$), within the F plane ($J_{F}>0$), between the AF and F planes ($J_{AF-F}>0$), and nnn interactions between the two AF planes ($J_{AF-AF}>0$). It is assumed that the F plane features an easy-axis single ion anisotropy ($A > 0$). Hereafter, to fix the energy scales in the respective layers, we set $J_{\parallel}=J_{AF}=-1$ and $J_{F}=1$.

The quantities of interest include the skyrmion chirality $\kappa$ - the discretization of a continuum topological charge~\cite{berg1981definition}. It provides information about the number and the nature of topological configurations present in the system and serves as the order parameter for the SkX phase. In each AF layer it can be calculated as



\begin{equation}
\kappa = \frac{\langle K \rangle}{N} = \frac{1}{4\pi N} \Big\langle \Big |\sum_i \Big( \kappa^{12}_{i} + \kappa^{34}_{i} \Big) \Big |\Big\rangle,
\label{chiral}
\end{equation}

where $\langle \hdots \rangle$ denotes the thermal average, $N=L^2$ is the number of spins in the layer, and $\kappa^{ab}_{i} = \vec{S_i}\cdot[\vec{S_a}\times\vec{S_b}]$ is the chirality of the triangular plaquette consisting of the central spin $\vec{S_i}$ and its neighboring spins $\vec{S_a}$ and $\vec{S_b}$, taken in the counter-clockwise direction~\cite{berg1981definition}. 

\begin{figure}[t!]
\centering
\subfigure{\includegraphics[scale=0.25,clip]{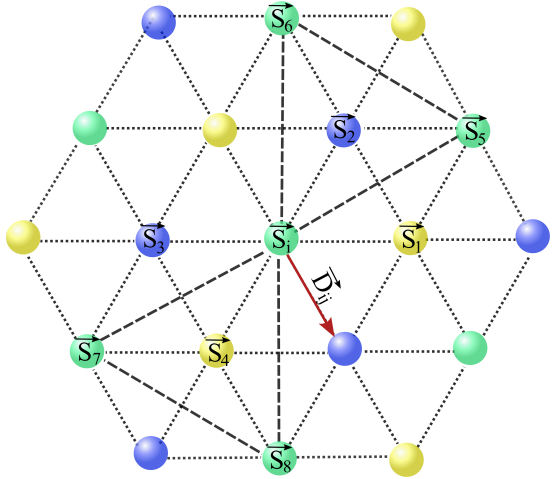}}
\caption{Three-sublattice decomposition of spins on a triangular lattice (circles of different colors), where $\vec{S_i}$ is the central spin, $\vec{S_1},\hdots,\vec{S_4}$ and $\vec{S_5},\hdots,\vec{S_8}$ are the spins involved in calculation of the local chirality and the skyrmion number, respectively, and $\vec{D}_{ij}$ represents the DMI vector.}
\label{fig:latt}
\end{figure}

Alternatively, one can also evaluate the skyrmion number $q$, which provides similar information, as

\begin{equation}
q = \frac{\langle Q \rangle}{N_s} = \frac{1}{4\pi N_s} \Big\langle\Big|\sum_i \Big( A^{56}_{i} \sign(\kappa^{56}_{i}) +A^{78}_{i} \sign( \kappa^{78}_{i}) \Big)\Big|\Big\rangle,
\label{sknum}
\end{equation}

where $A^{ab}_{i} = ||(\vec{S_a} - \vec{S_i})\times(\vec{S_b} - \vec{S_i})||/2$ is the area of the triangle spanned by those spins. The chirality is calculated for the whole AF layer and the summation runs through all the spins with $\{\vec{S_a}, \vec{S_b}\}$ corresponding to  $\{\vec{S_1}, \vec{S_2}\}$ and $\{\vec{S_3}, \vec{S_4}\}$ (see Fig.~\ref{fig:latt}), whereas the skyrmion number is calculated for each of the three sublattices of the layer, hence $N_s$ represents the number of sites in each of the sublattices, $N_s = L^2/3$, and the triangular plaquette for the local quantities is formed by the neighboring spins of the given sublattice $\{\vec{S_i}, \vec{S_5}, \vec{S_6}\}$ and $\{\vec{S_i}, \vec{S_7},\vec{S_8}\}$, as shown in Fig.~\ref{fig:latt}. Spins in both cases are taken in a counter-clockwise fashion to keep the sign in accordance with the rules in ~\cite{berg1981definition}.

The specific heat in the AF plane can be calculated as

\begin{equation}
c = \frac{\langle \mathcal{H}_{AF}^2\rangle  - \langle \mathcal{H}_{AF} \rangle^2 }{NT^2},
\label{spec_heat}
\end{equation}
where $\mathcal{H}_{AF}$ refers to the Hamiltonian pertaining to the AF plane. The above quantities are averaged over all AF layers in the respective systems.


To accelerate the simulations we employed the  hybrid MC method, combining the stochastic Metropolis algorithm with the deterministic energy-preserving over-relaxation method~\cite{creutz1987overrelaxation}. The latter helps decorrelate the system and leads to faster relaxation. Each layer consists of $N = L^2$ sites with the periodic boundary conditions applied in the plane. After testing several values of the lattice size we chose $L = 48$, beyond which finite-size effects are already negligible and larger values give similar results. For thermal averaging we used up to $5 \times 10^5$ MC steps (MCS) after discarding another up to $10^6$ MCS for equilibration. The simulations were massively parallelized on General Purpose Graphical Processing Units in CUDA environment. 


\section{Results}
\subsection{AF multilayer}
\begin{figure}[t!]
\centering
\subfigure{\includegraphics[scale=0.55,clip]{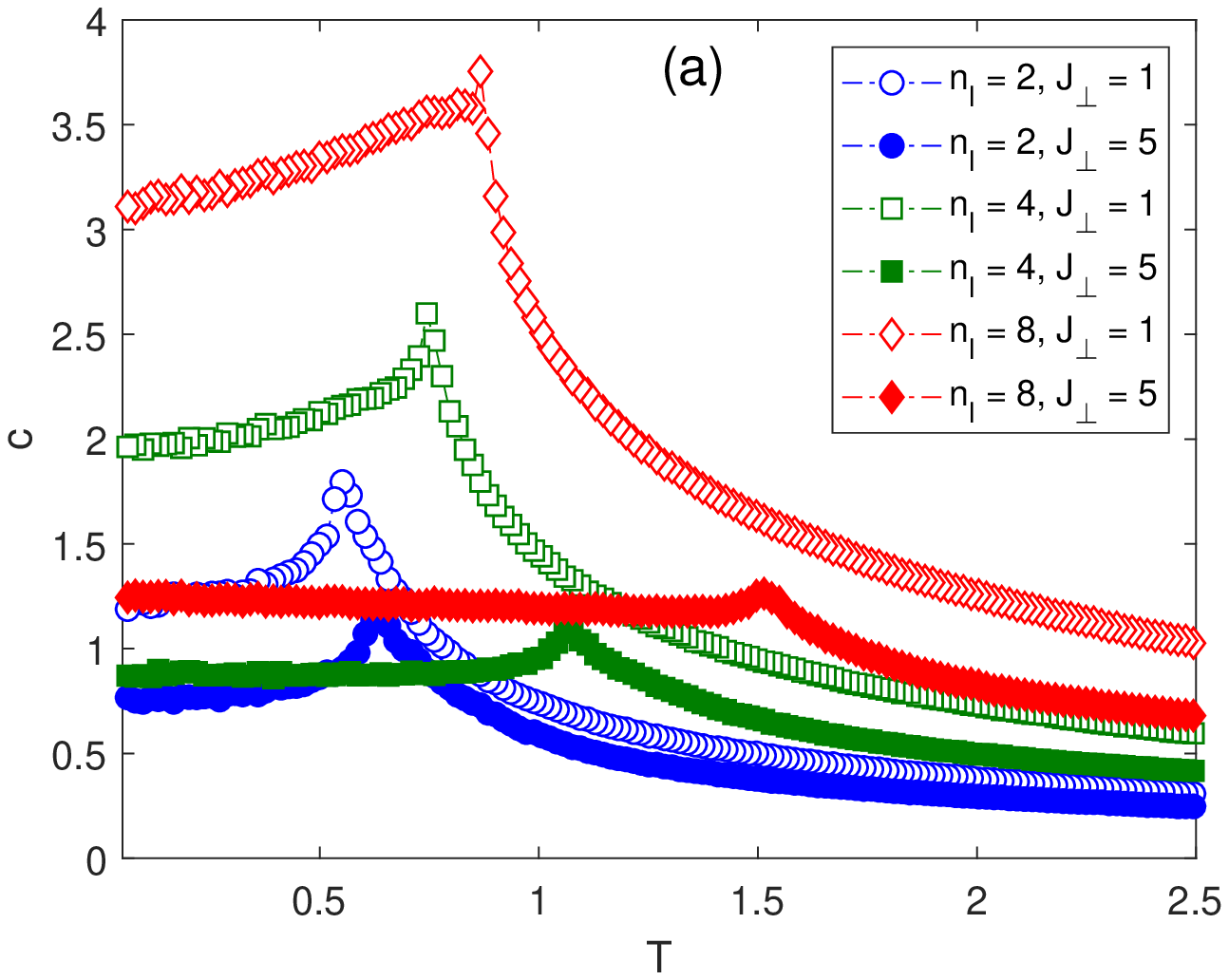}\label{fig:c-T}} \\
\subfigure{\includegraphics[scale=0.55,clip]{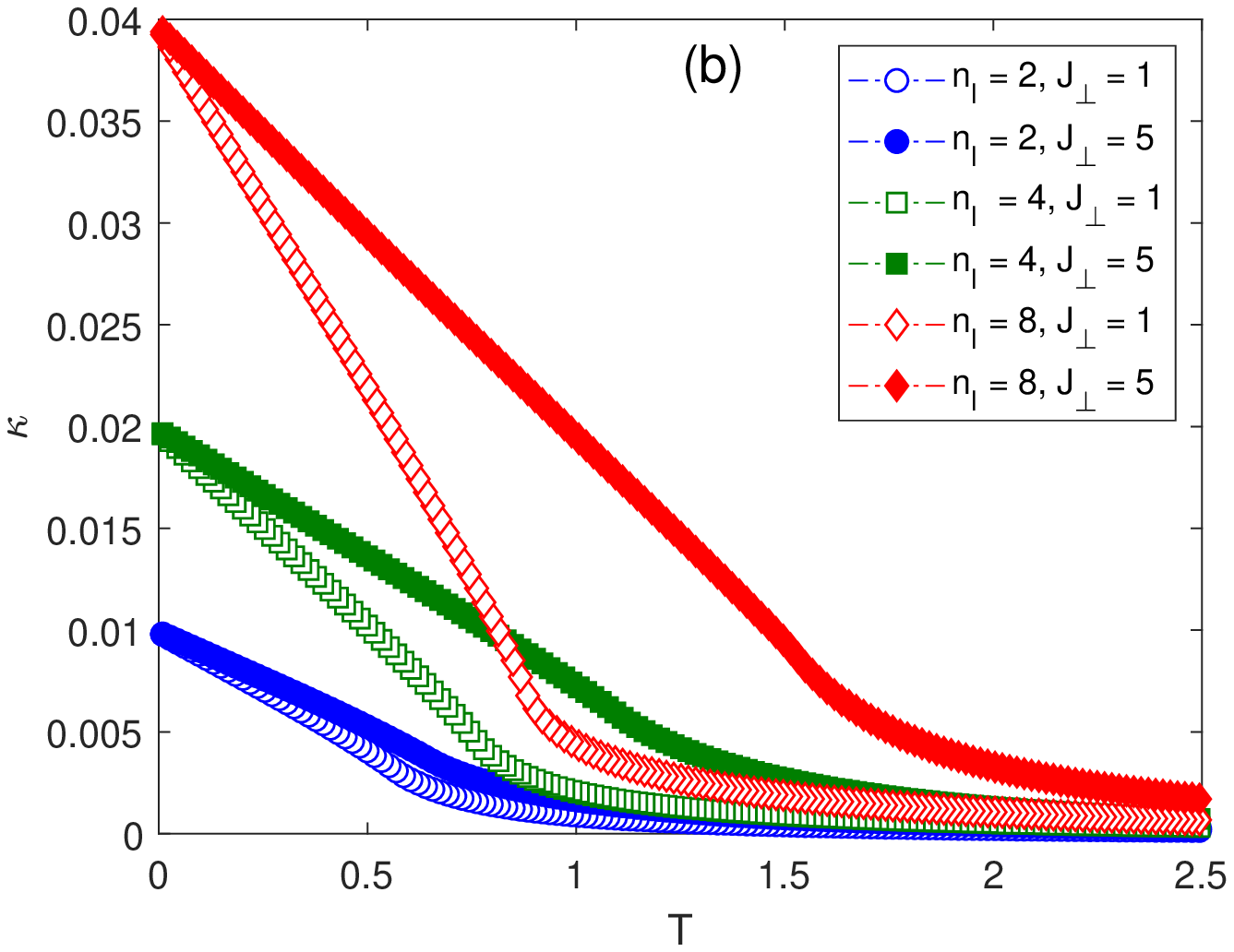}\label{fig:kappa-T}}
\subfigure{\includegraphics[scale=0.55,clip]{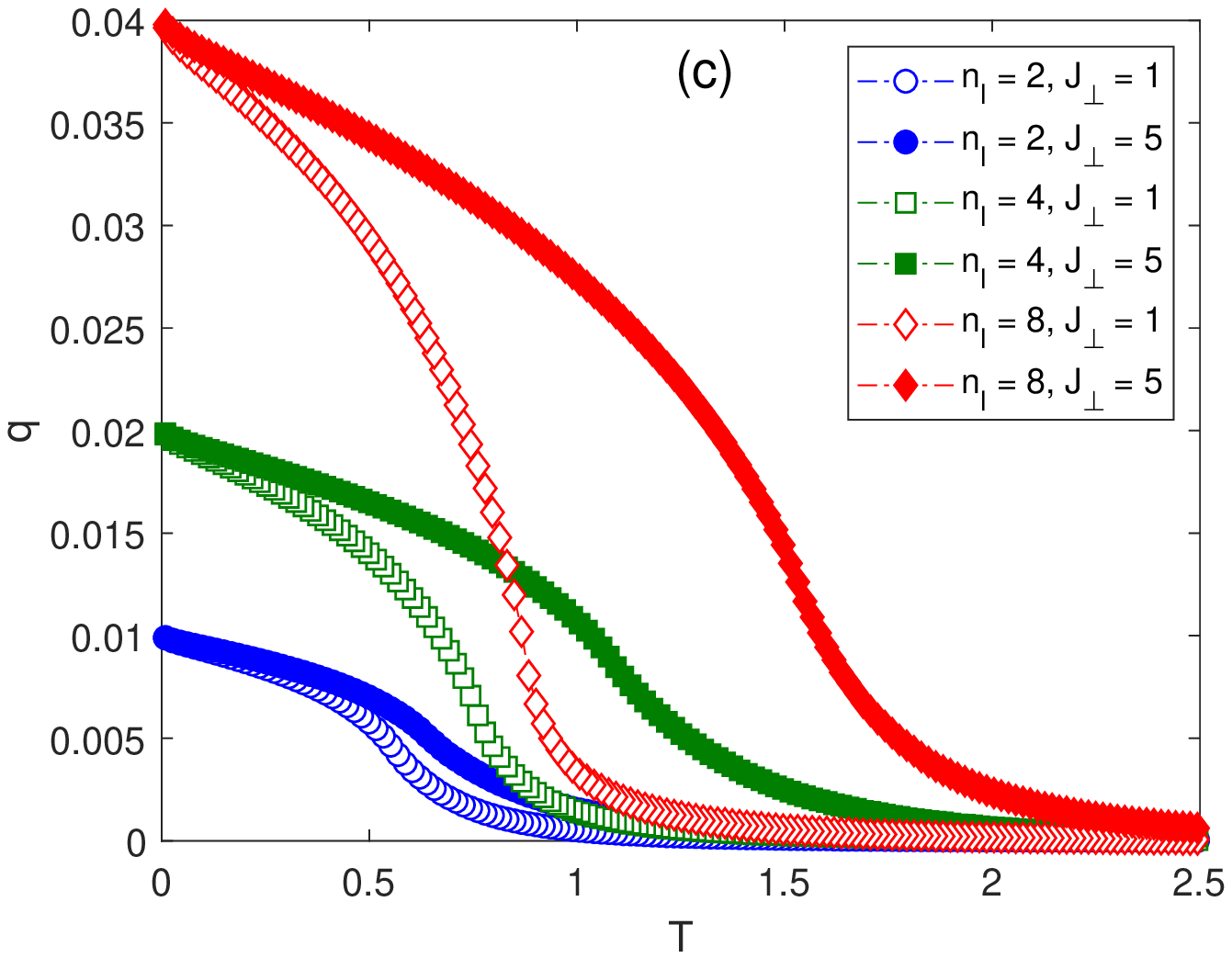}\label{fig:q1-T}}
\caption{Temperature dependencies of (a) the specific heat, (b) the chirality and (c) the skyrmion number, for $h=3$ and various values of the multilayer thickness, $n_l$, and the interlayer coupling, $J_{\perp}$.}
\label{fig:multi_x-T}
\end{figure}

Compared to the single-layer model, the multilayer system involves two additional parameters - the number of layers $n_l$ and the interlayer coupling $J_{\perp}$, that can play an important role in controlling the SkX phase extent and stability. To focus on the effects of these parameters on the transition temperature $T_{c}^{SkX}$ between the SkX and the paramagnetic (P) phase we restrict the remaining parameters to the region of the SkX phase emergence in a single layer and study temperature variations of various relevant quantities. In particular, we fix the DMI intensity to $D=0.5$ and in most cases the external magnetic field to $h=3$. The quantities relevant for the SkX phase identification include the specific heat, the chirality, and the skyrmion number. These are plotted in Fig.~\ref{fig:multi_x-T} as functions of temperature for various values of the parameters $n_l=2,4$ and $8$ and $J_{\perp}=1$ and $5$. 

In Fig.~\ref{fig:c-T} one can notice in each specific heat curve a single sharp peak indicating the presence of a phase transition. The remaining panels~\ref{fig:kappa-T} and~\ref{fig:q1-T}, showing respectively the chirality and the skyrmion number - the order parameters for the SkX phase - tell us that the phase transition is to the SkX phase appearing below $T_{c}^{SkX}$. To see the effect of the individual parameters on the SkX-P transition temperature $T_{c}^{SkX}$ it is useful to focus on the curves with one parameter fixed and the other one varying. Apparently, for either value of $J_{\perp}$, as the number of layers $n_l$ increases the peaks of the specific heat curves increase and shift to higher temperatures. Namely, from $T_{c}^{SkX} \approx 0.55$ for $n_l=2$ up to $T_{c}^{SkX} \approx 0.87$ for $n_l=8$ in the case of $J_{\perp}=1$ and from $T_{c}^{SkX} \approx 0.64$ for $n_l=2$ up to $T_{c}^{SkX} \approx 1.51$ for $n_l=8$ in the case of $J_{\perp}=5$. By comparing the $T_{c}^{SkX}$ values for the same number of layers and different $J_{\perp}$ one can make a similar conclusion: the increasing interlayer interaction leads to an increase in the $T_{c}^{SkX}$ transition temperature\footnote{The value of $T_{c}^{SkX}$ actually gives us a pseudo-transition temperature, corresponding to the chosen lattice size $L$.}.

\begin{figure}[t!]
\centering
\subfigure{\includegraphics[scale=0.55,clip]{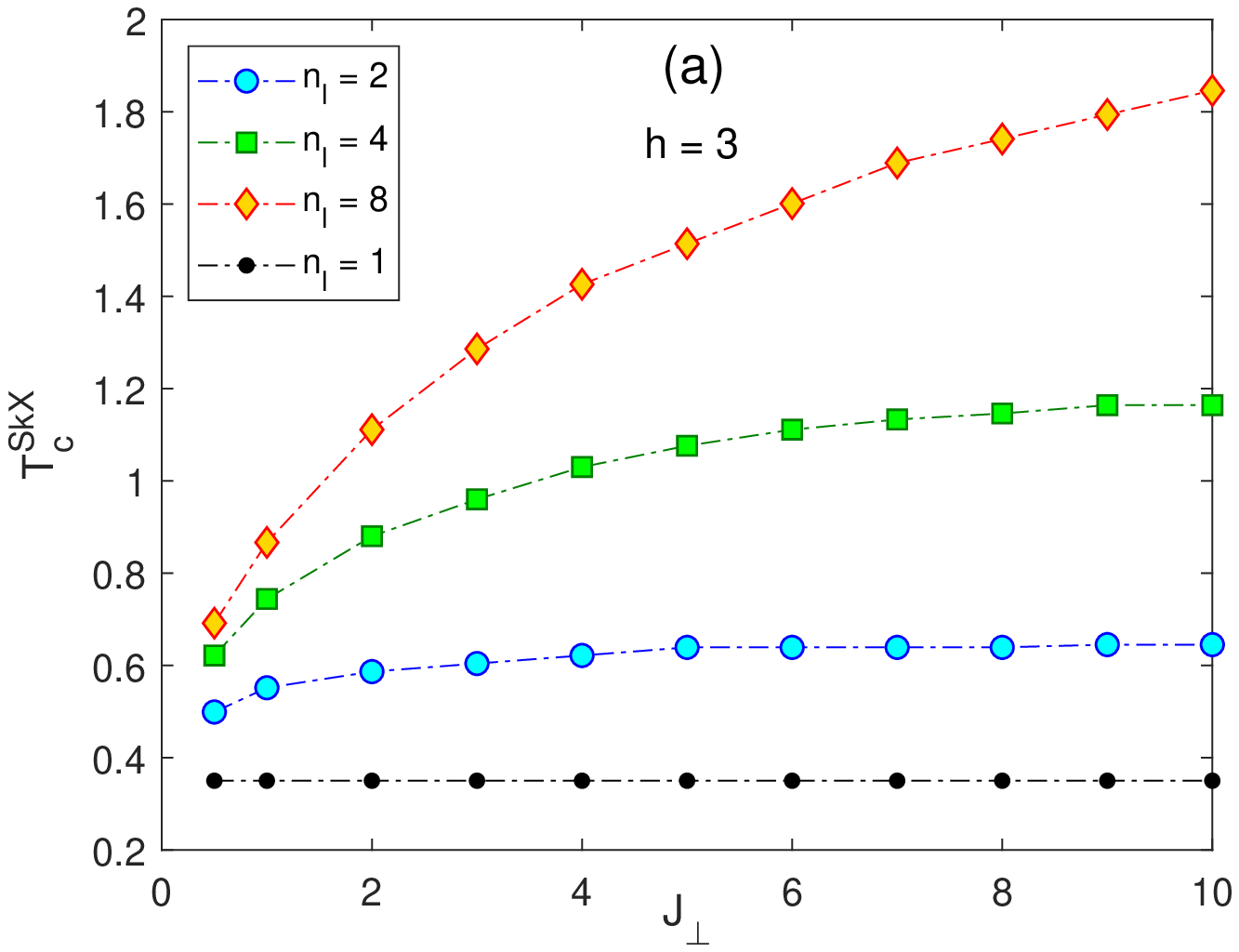}\label{fig:Tc-Ji_h3}}
\subfigure{\includegraphics[scale=0.55,clip]{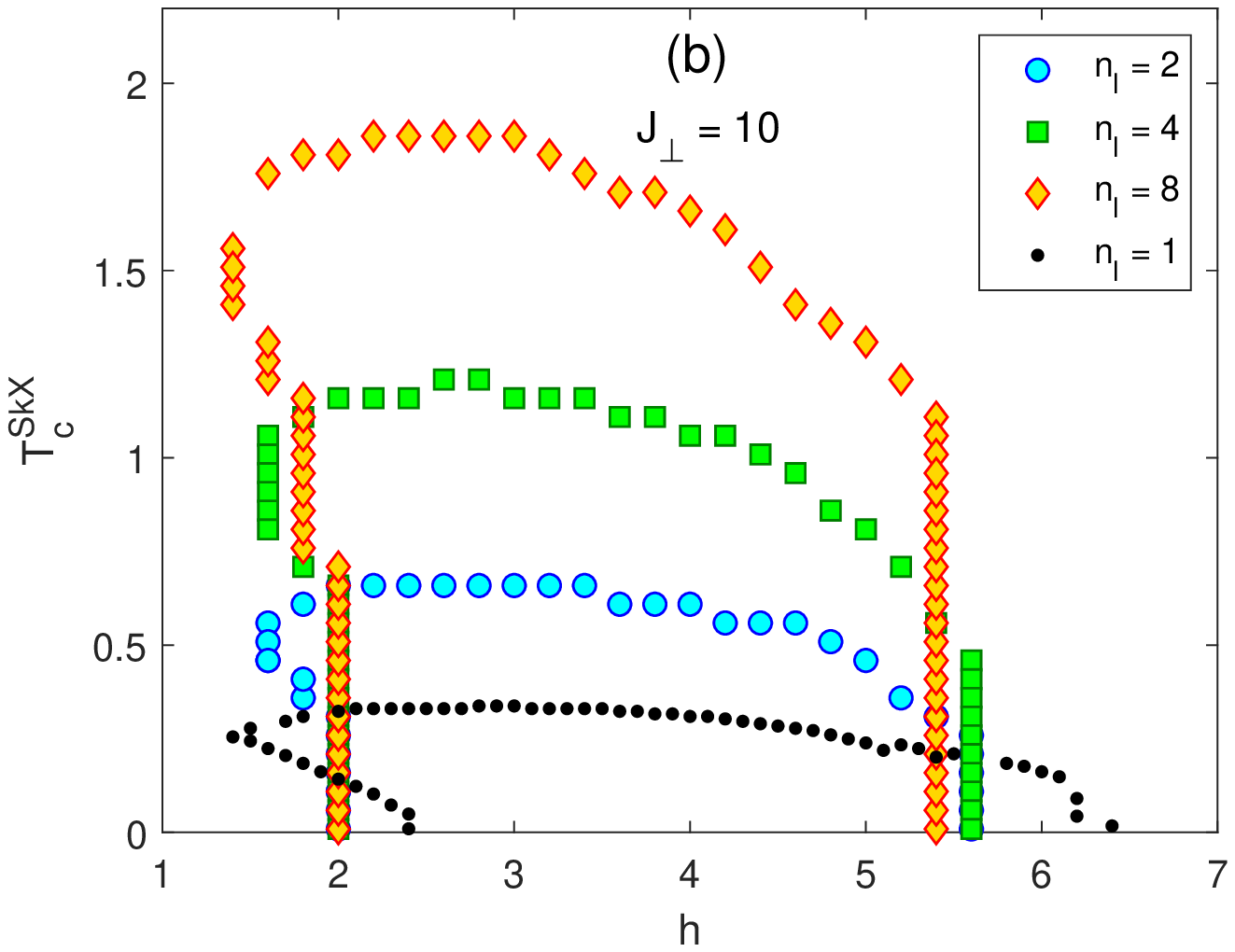}\label{fig:Tc-h_Ji10}}
\caption{Transition temperatures to the SkX phase, $T_{c}^{SkX}$, as functions of (a) the interlayer coupling at $h=3$ and (b) the magnetic field at $J_{\perp}=10$, for different values of $n_l$. The black dots indicate the transition temperatures for a single layer.}
\label{fig:multi_PD}
\end{figure}

These findings are summarized in Fig.~\ref{fig:multi_PD}, which shows systematic variations of the $T_{c}^{SkX}$ transition temperature with $J_{\perp}$ for $h=3$ (Fig.~\ref{fig:Tc-Ji_h3}) and with $h$ for $J_{\perp}=10$ (Fig.~\ref{fig:Tc-h_Ji10}), and different values of $n_l$. From the curves in Fig.~\ref{fig:Tc-Ji_h3} it is apparent that the increase of $T_{c}^{SkX}$ is more pronounced at smaller $J_{\perp}$ and then the curves seem to converge to some maximum values, which stay constant after further increase of $J_{\perp}$. However, with the increasing number of layers the increase of $T_{c}^{SkX}$ can be observed over larger and larger intervals of $J_{\perp}$. For example, for $n_l=8$ the increase seems to continue even beyond $J_{\perp}=10$, at which the transition temperature has reached $T_{c}^{SkX} \approx 1.85$. For reference, we also include the case of $n_l=1$, i.e. the transition temperature for the single layer, $T_{c}^{SkX} \approx 0.35$, at $h=3$. Thus, one can see that the transition temperature can be increased by more than five times in the multilayers composed of a sufficient number of layers coupled with sufficiently large interlayer interaction.

\begin{figure}[t!]
\centering
\subfigure{\includegraphics[scale=0.45,clip]{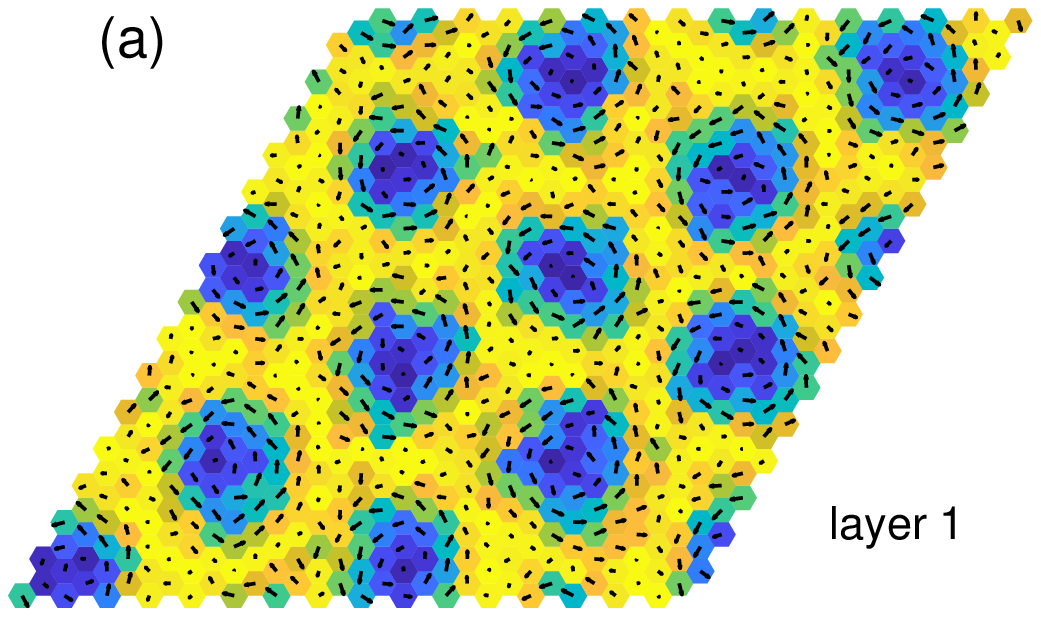}\label{fig:snap_T=0_2015_Jint=1_nl=8_L1}}
\subfigure{\includegraphics[scale=0.45,clip]{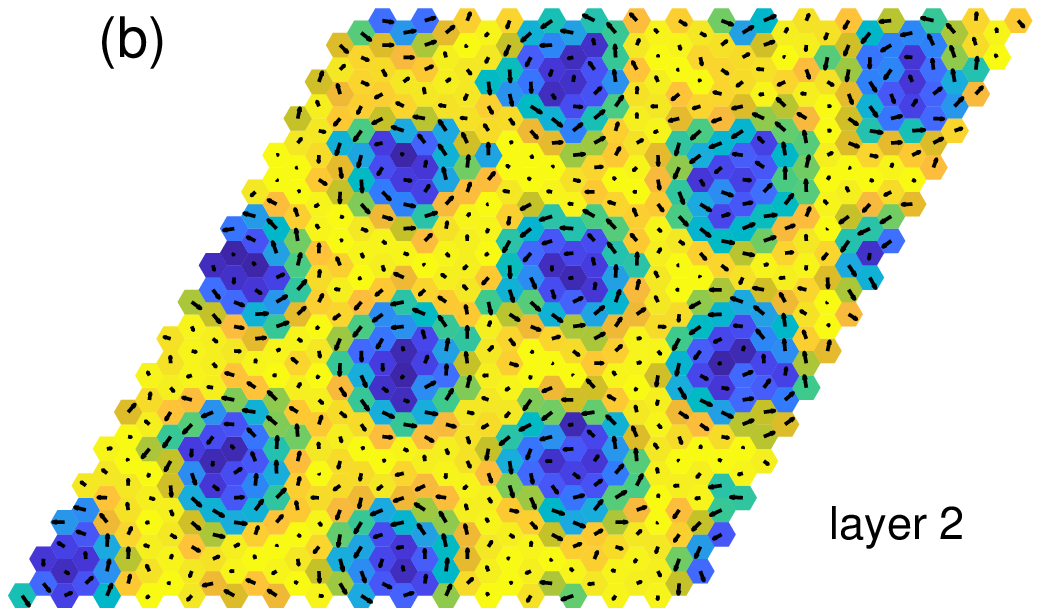}\label{fig:snap_T=0_2015_Jint=1_nl=8_L2}}
\subfigure{\includegraphics[scale=0.45,clip]{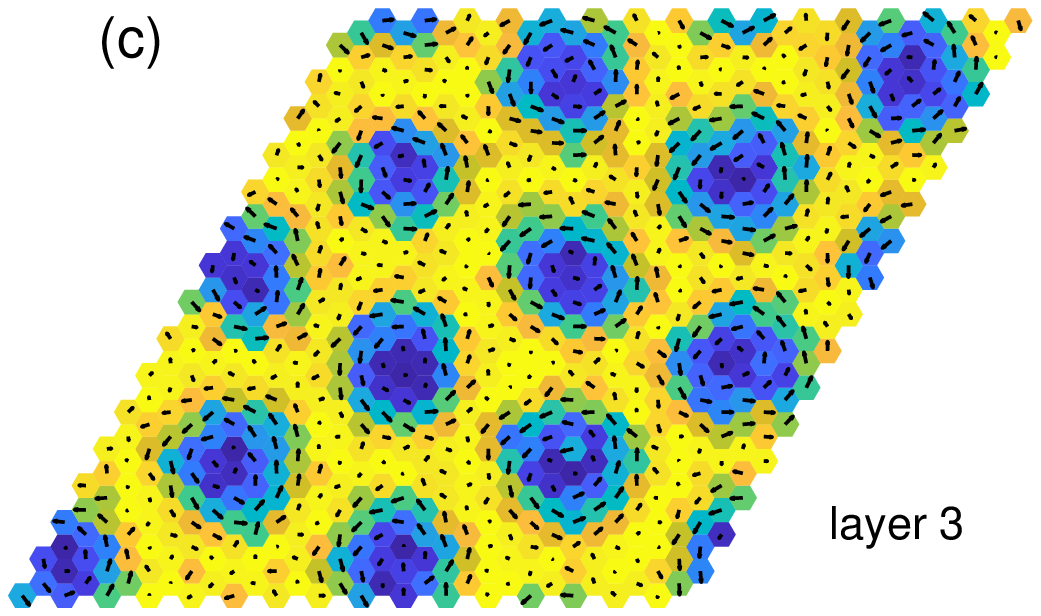}\label{fig:snap_T=0_2015_Jint=1_nl=8_L3}}\\
\subfigure{\includegraphics[scale=0.45,clip]{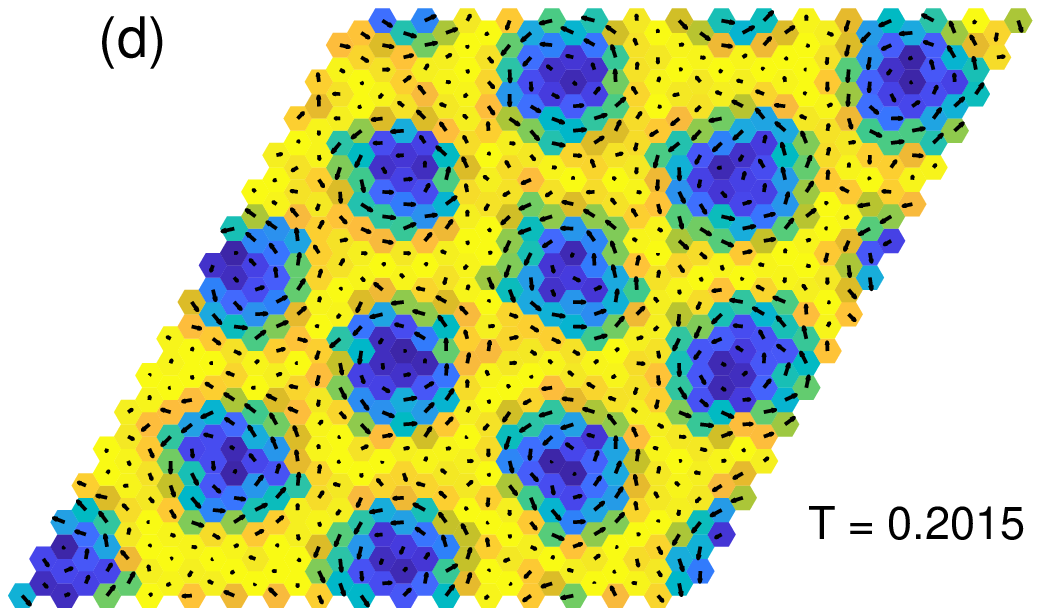}\label{fig:snap_T=0_2015_Jint=1_nl=8_L4}}
\subfigure{\includegraphics[scale=0.45,clip]{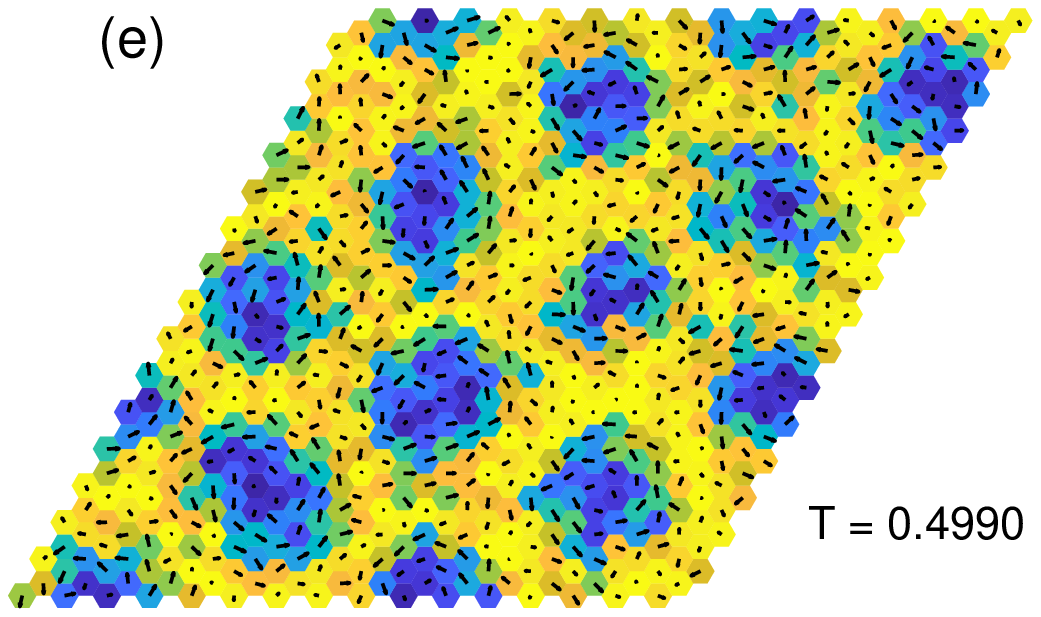}\label{fig:snap_T=0_4990_Jint=1_nl=8_L4}}
\subfigure{\includegraphics[scale=0.45,clip]{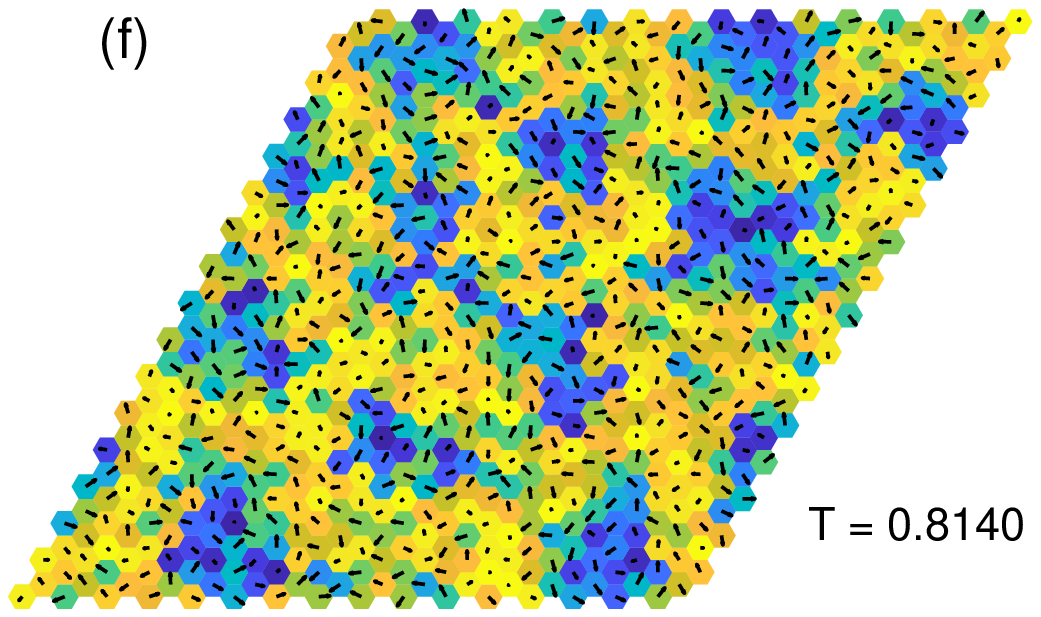}\label{fig:snap_T=0_8140_Jint=1_nl=8_L4}}\\
\subfigure{\includegraphics[scale=0.45,clip]{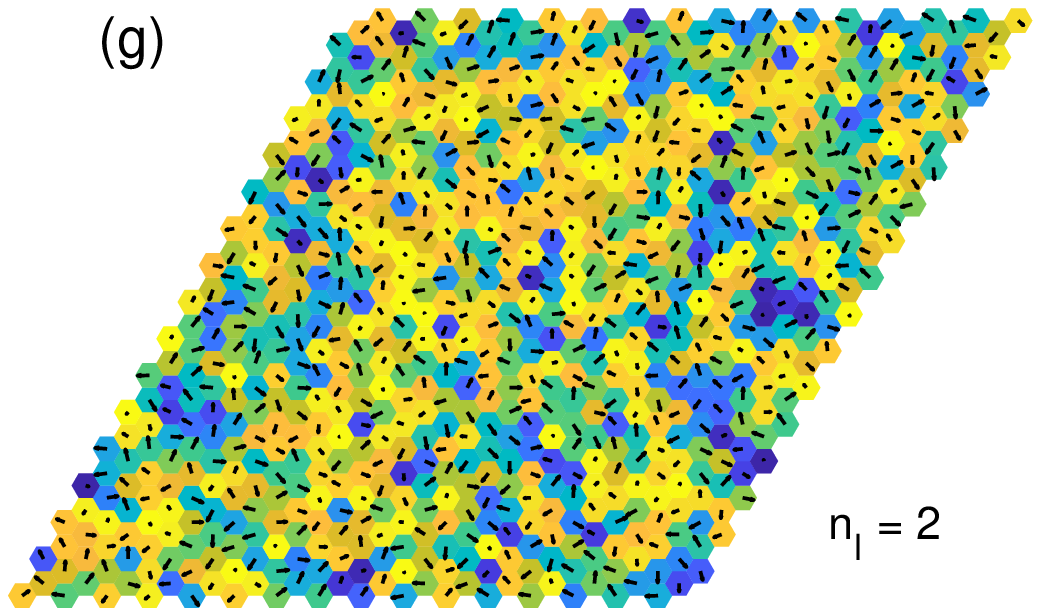}\label{fig:snap_T=1_0065_Jint=10_nl=2}}
\subfigure{\includegraphics[scale=0.45,clip]{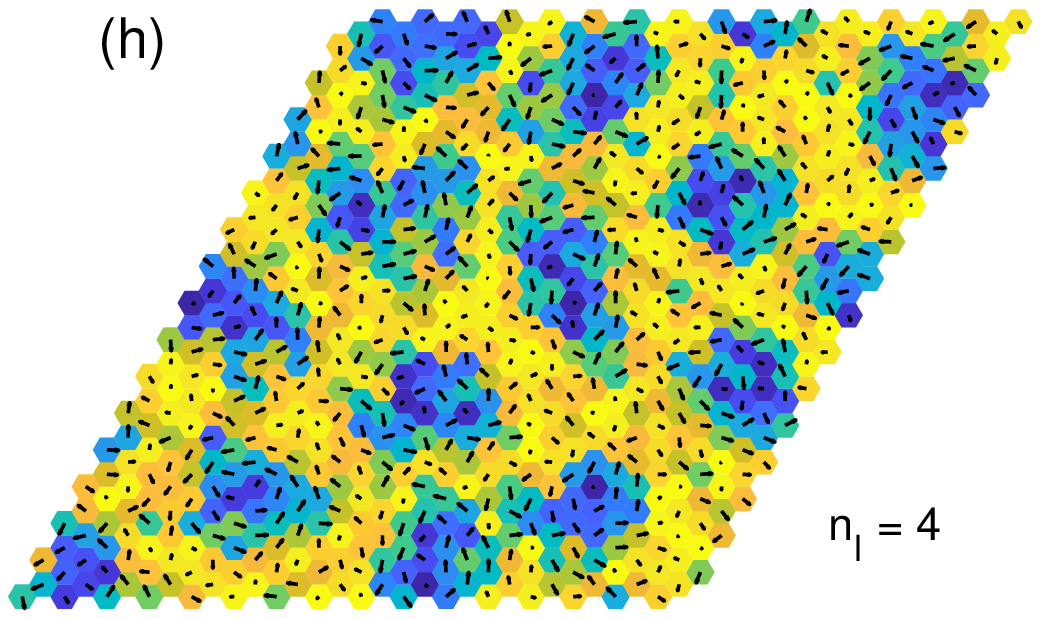}\label{fig:snap_T=1_0065_Jint=10_nl=4}}
\subfigure{\includegraphics[scale=0.45,clip]{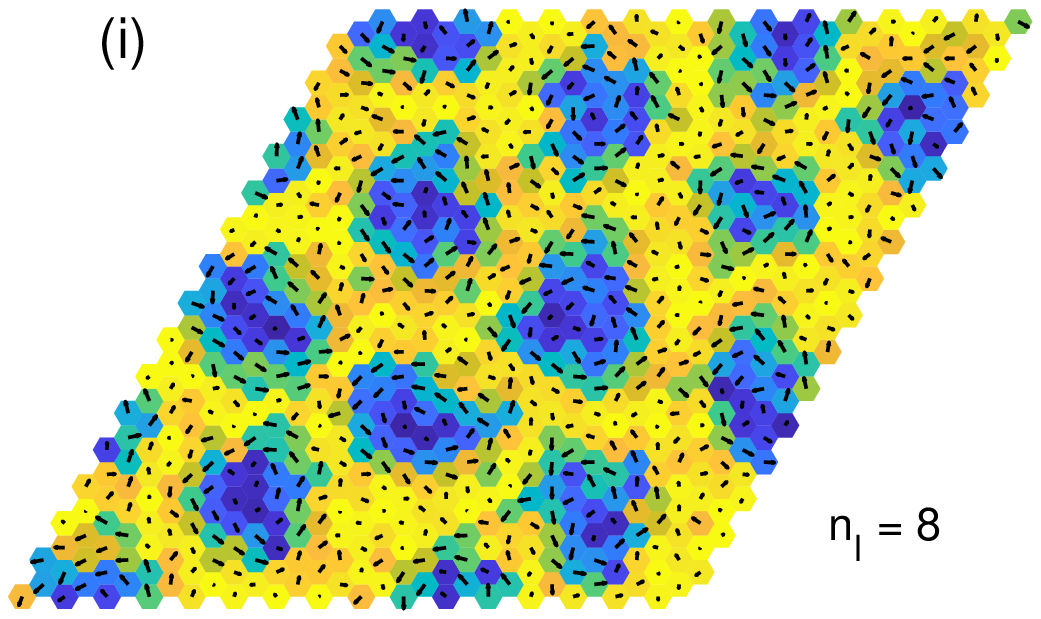}\label{fig:snap_T=1_0065_Jint=10_nl=8}}
\caption{Sublattice spin snapshots visually demonstrating the SkX phase stability in the multilayer system for $D=0.5$ and $h=3$ with the increasing (a-c) layer labels for $n_l=8$, $J_{\perp}=1$ and $T=0.2015$, (d-f) temperature for $n_l=8$, $J_{\perp}=1$ and $l=4$, and (g-i) number of layers for $J_{\perp}=10$ and $T=1.0065$.}
\label{fig:multi_snaps}
\end{figure}

Then the question is the stability of the SkX phase under different conditions, such as in the bulk and on the surface of the multilayer structure or persistence of the skyrmion lattice structure at increased temperatures. It is reasonable to assume that the SkX structure in the surface layers with the free boundary conditions on one side is more susceptible to disruption due to different perturbations than the bulk layers sandwiched between other layers. The top row in Fig.~\ref{fig:multi_snaps} shows the SkX phase real space sublattice spin configurations in the first (surface), second, and third layers of the multilayer consisting of $n_l=8$ layers. A closer look indeed reveals a slightly increasing irregularity of the SkX patterns as we progress from the interior to the surface but the differences are barely noticeable and the SkX structure is clearly present even in the surface layer.

The middle and bottom rows in Fig.~\ref{fig:multi_snaps} respectively demonstrate the effects of the increasing temperature and the multilayer thickness on the stability of the skyrmion lattice structure. The middle row presents the spin snapshots in the $n_l=8$ multilayer (represented by the bulk layer number $l = 4$) with $J_{\perp}=1$ taken within the SkX phase at the gradually increasing temperature. As expected, thermal fluctuation gradually destroy the well defined SkX texture observable at low temperatures. From Fig.~\ref{fig:Tc-Ji_h3} we expect a complete destruction of the SkX structure and the transition to the paramagnetic phase at $T_{c}^{SkX} \approx 0.86$. The last snapshot (Fig.~\ref{fig:snap_T=0_8140_Jint=1_nl=8_L4}), taken relatively close to the transition point at $T = 0.814$, shows already a great degree of thermal fluctuations, nevertheless, the SkX texture can still be discerned. Finally, the bottom row demonstrates the effect of gradual doubling of the number of the layers in the system with $J_{\perp}=10$ at $T \approx 1$. In particular, we can first notice the phase transition from the paramagnetic state observed in the bilayer system (Fig.~\ref{fig:snap_T=1_0065_Jint=10_nl=2}) to the SkX state for $n_l=4$ (Fig.~\ref{fig:snap_T=1_0065_Jint=10_nl=4}) and further improvement of the skyrmion contours for $n_l=8$ (Fig.~\ref{fig:snap_T=1_0065_Jint=10_nl=8}).

\subsection{AF-F-AF sandwich}
Above we showed that ferromagnetic stacking of the AF layers on top of each other leads to an extension of the SkX phase to higher temperatures, which is particularly important in technological applications. Equally important is to facilitate its emergence at accessible, i.e. sufficiently small or zero, external magnetic fields. In our previous paper~\cite{mohylna2022road} we have demonstrated that the SkX phase in the present model can be realized at very small or even zero external magnetic fields. It can be achieved by its coupling to another sufficiently hard F layer, which can serve as a source of an auxiliary effective field acting on the AF plane alongside the external magnetic field. Below we show that by creating a superstructure, composed of two coupled AF layers and one F layer sandwiched in between, one can achieve that the SkX phase is realized at significantly elevated temperatures and very weak or zero magnetic field.

\begin{figure}[t!]
\centering
\subfigure{\includegraphics[scale=0.55,clip]{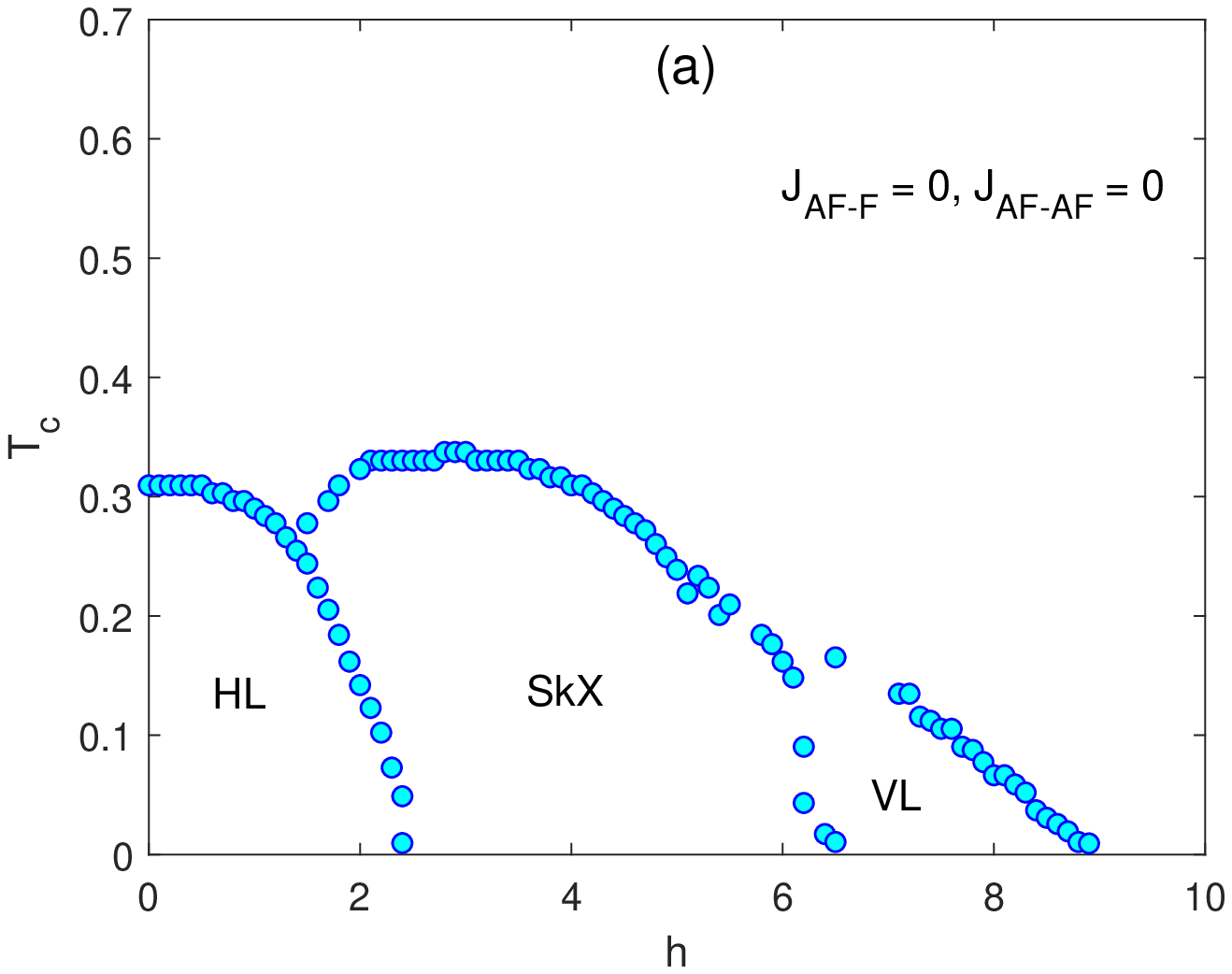}\label{fig:sand_T-h_Ji0}}
\subfigure{\includegraphics[scale=0.55,clip]{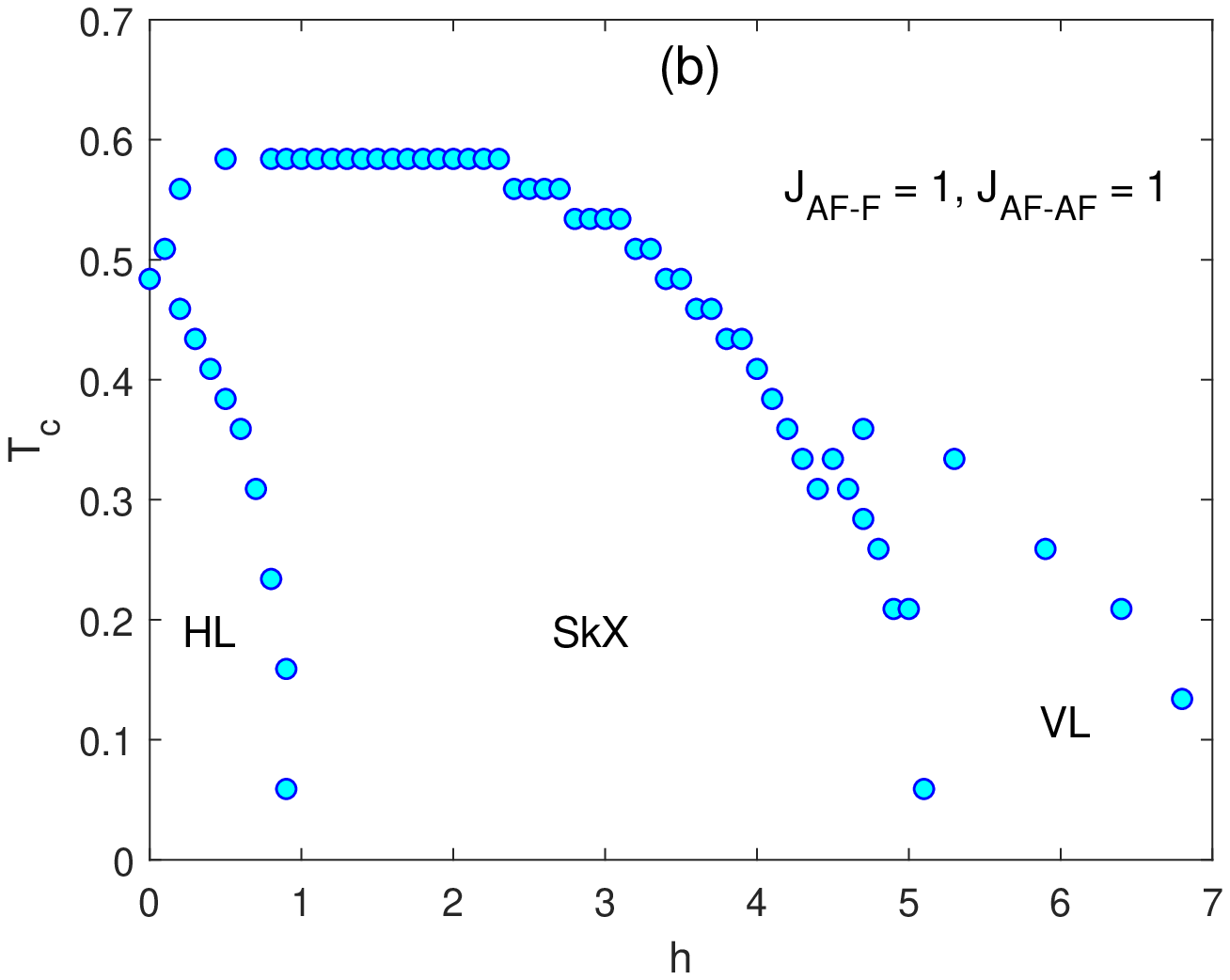}\label{fig:sand_T-h_Ji1}}
\subfigure{\includegraphics[scale=0.55,clip]{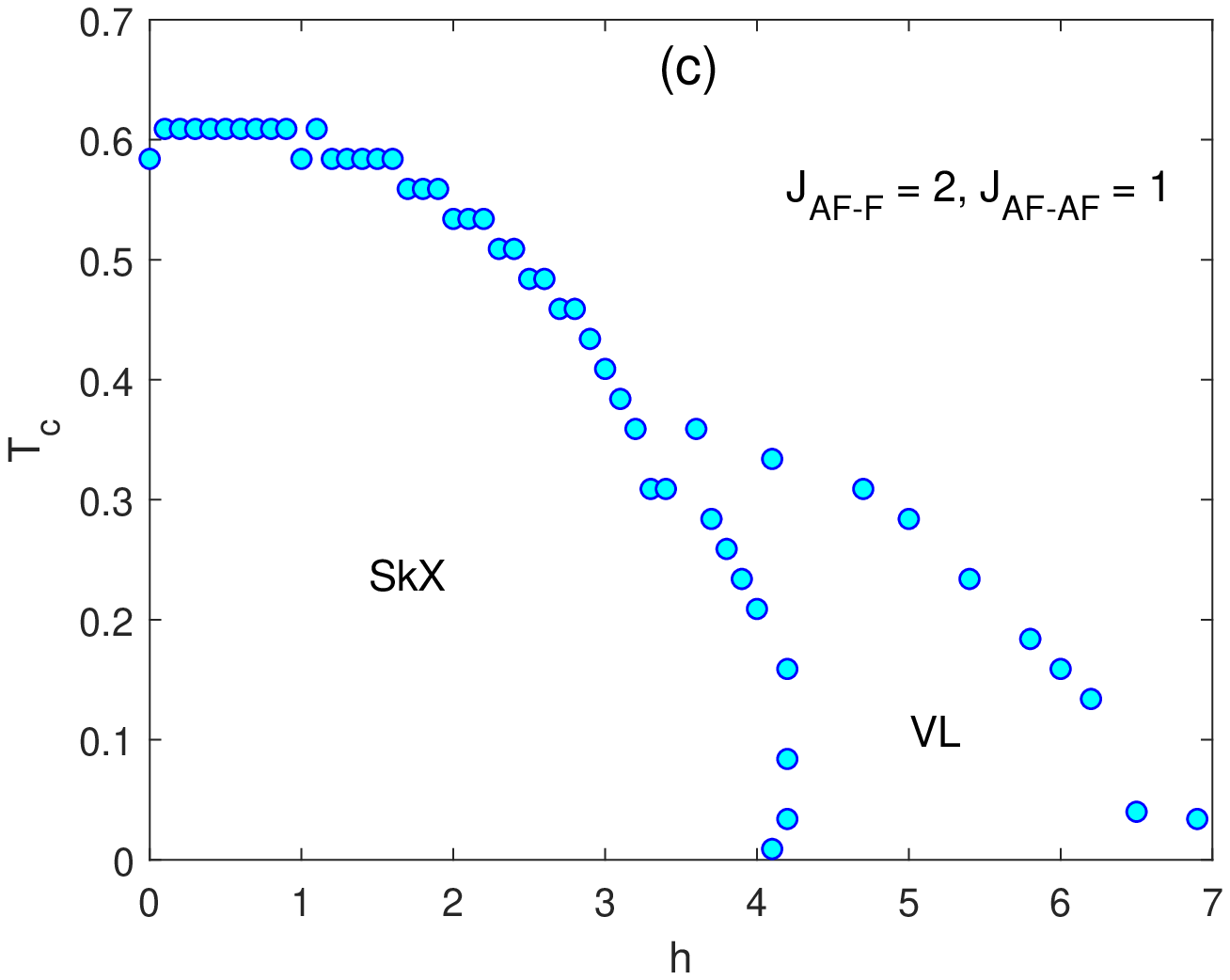}\label{fig:sand_T-h_Ji2}}
\caption{Phase diagrams of the AF layers, as functions of the magnetic field, for $D=0.5$, $A=1$ and (a) $J_{AF-AF}=0$, $J_{AF-F}=0$ (single layer), (b) $J_{AF-AF}=1$, $J_{AF-F}=1$ and (c) $J_{AF-AF}=1$, $J_{AF-F}=2$. HL, SkX and VL denote the helical, skyrmion and V-like phases, respectively. }
\label{fig:sand_PD}
\end{figure}

Figure~\ref{fig:sand_T-h_Ji0} shows the entire phase diagram for the single AF layer, i.e. with the sandwich structure parameters set to $J_{AF-AF}=J_{AF-F}=0$. The SkX phase extends within the field range $2.4 \lesssim h \lesssim 6.5$ for low temperatures and the temperatures up to $T \approx 0.35$. The model also displays a helical (HL) phase at smaller and a V-like (VL) phase at higher fields~\cite{rosales2015three,mohylna2021formation}. The effect of switching on both $J_{AF-AF}$ and $J_{AF-F}$ interactions on the SkX phase is apparent from Fig.~\ref{fig:sand_T-h_Ji1}. For $J_{AF-AF}=J_{AF-F}=1$ the temperature extent of the SkX phase almost doubles and the field window shifts to smaller values by $\Delta h \approx 1$. Further decrease of the external field can be achieved by strengthening the effective field imposed by the F layer, i.e. by increasing $J_{AF-F}$. Figure~\ref{fig:sand_T-h_Ji2} indicates that for $J_{AF-F}=2$ the HL phase is completely replaced by the SkX phase. Thus, the latter can be realized even at zero field up to the temperatures almost twice as high as in the single layer.

\section{Conclusion}
Finally, we have explored some possibilities of bringing the SkX phase realization in the AF Heisenberg antiferromagnet to the conditions more plausible for experimental observations and technological applications. In particular, we demonstrated that by stacking of multiple AF layers on top of each other and coupling them with a sufficiently strong F interaction we can achieve that the SkX phase will persist to much higher temperatures than in the single layer. For example, within the parameters ranges used in our simulations the SkX-P transition temperature in the multilayer consisting of eight AF layers coupled by sufficiently strong interaction was more than five times higher, compared to the single layer. Our results also indicated that by using the structures with more layers and larger interlayer interactions the SkX phase could be brought to even much higher temperatures. Thus, by tailoring multilayer structures it might be possible to design skyrmion-based technological devices that can operate at room or even higher temperatures.

Another challenge is a very strong magnetic field that is required for the stabilization of the skyrmion lattice phase in the present model in a single layer. For example, for Fe/MoS$_2$ - a potential experimental realization - this field could be on the order of 20 T~\cite{fang2021spirals}. We showed that by coupling the AF layer to the reference F layer, this field can be drastically reduced or even completely eliminated. Thus, by combining the two above approaches one can design a heterostructure that would display the SkX phase at relatively high (room) temperatures and zero magnetic field, which is the environment favorable for technological applications.

\section*{CRediT authorship contribution statement}
\textbf{Mariia Mohylna}: Methodology, Software, Data curation, Visualization, Formal analysis, Writing – original draft. \textbf{Vitalii Tkachenko}: Software, Data curation, Formal analysis. \textbf{Milan \v{Z}ukovi\v{c}}: Conceptualization, Investigation, Validation, Writing – original draft, Writing – review \& editing, Supervision, Resources, Project administration.

\section*{Declaration of competing interest}
The authors declare no competing financial interest.

\section*{Data availability}
No data was used for the research described in the article.

\section*{Acknowledgment}
This work was supported by the grants of the Slovak Research and Development Agency (Grant No. APVV-20-0150) and the Scientific Grant Agency of Ministry of Education of Slovak Republic (Grant No. 1/0695/23).

\bibliographystyle{elsarticle-num}
\bibliography{multi}

\end{document}